\documentclass[prb,preprint,preprintnumbers,amsmath,amssymb]{revtex4}
\usepackage{graphicx}% Include figure files
\usepackage{amsmath}
\usepackage{amsbsy}
\usepackage{dcolumn}% Align table columns on decimal point
\usepackage{bm}% bold math

\begin{document}

\title{\flushleft{\Huge\sf Nonequilibrium Singlet-Triplet Kondo Effect in Carbon Nanotubes}}

\author{\begin{flushleft}{\sf J. PAASKE$^{1\ast}$, A. ROSCH$^{2}$, P. W\"{O}LFLE$^{3}$,
N. MASON$^{4,5}$, C. M. MARCUS$^{5}$ AND J.
NYG{\AA}RD,$^{1}$}\end{flushleft}} \affiliation{\begin{flushleft}
\footnotesize\normalfont\mbox{\sf $^{1}$The Niels Bohr Institute \&
The Nano-Science Center, University of Copenhagen,
%Universitetsparken 5,
DK-2100 Copenhagen, Denmark}\\
\footnotesize\normalfont\mbox{\sf $^{2}$ Institut f\"{u}r
Theoretische
Physik, Universit\"{a}t zu K\"{o}ln, 50937 K\"{o}ln, Germany}\\
\footnotesize\normalfont\mbox{\sf $^{3}$ Institut f\"{u}r Theorie
der Kondensierten
Materie, Universit\"{a}t Karlsruhe, 76128 Karlsruhe, Germany}\\
\footnotesize\normalfont\mbox{\sf $^{4}$
Department of Physics, University of Illinois at Urbana Champaign, Urbana IL 61801-3080, USA} \\
\footnotesize\normalfont\mbox{\sf $^{5}$
Department of Physics, Harvard University, Cambridge MA 0213, USA} \\
\footnotesize\normalfont\mbox{\sf $^\ast$e-mail:
paaske@fys.ku.dk.}\end{flushleft}}
\date{\sf\today}

\maketitle

%%%%%%%%%%%%%%%%%%%%%%%%%%%%%%%%%%%%%%%%%%%%%%%%%%%%%%%%%%%%%%%%%%%%%%%%%%%%%%%%%
%%%%%%%%%%%%%%%%%%%%%%%%%%%% Introductory paragraph %%%%%%%%%%%%%%%%%%%%%%%%%%%%%
%%%%%%%%%%%%%%%%%%%%%%%%%%%%%%%%%%%%%%%%%%%%%%%%%%%%%%%%%%%%%%%%%%%%%%%%%%%%%%%%%

{\bf The Kondo-effect is a many-body phenomenon arising due to
conduction electrons scattering off a localized
spin~\cite{Hewson93}. Coherent spin-flip scattering off such a
quantum impurity correlates the conduction electrons and at low
temperature this leads to a {\it zero-bias} conductance
anomaly~\cite{Glazman88,Ng88}. This has become a common signature in
bias-spectroscopy of single-electron transistors, observed in GaAs
quantum
dots~\cite{Goldhaber98,Cronenwett98,vanderWiel00,Sasaki00,Kogan03,Zumbuhl04}
as well as in various single-molecule
transistors~\cite{Nygaard00,Park02,Liang02b,Yu04b,Babic04,Herrero05}.
While the zero-bias Kondo effect is well established it remains
uncertain to what extent Kondo correlations persist in
non-equilibrium situations where inelastic processes induce
decoherence. Here we report on a pronounced conductance peak
observed at finite bias-voltage in a carbon nanotube quantum dot in
the spin singlet ground state. We explain this {\it finite-bias}
conductance anomaly by a nonequilibrium Kondo-effect involving
excitations into a spin triplet state. Excellent agreement between
calculated and measured nonlinear conductance is obtained, thus
strongly supporting the correlated nature of this nonequilibrium
resonance.}

%%%%%%%%%%%%%%%%%%%%%%%%%%%%%%%%%%%%%%%%%%%%%%%%%%%%%%%%%%%%%%%%%%%%%%%%%%%%%%%%%
%%%%%%%%%%%%%%%%%%%%%%%%%%%%%% Further Introduction %%%%%%%%%%%%%%%%%%%%%%%%%%%%%
%%%%%%%%%%%%%%%%%%%%%%%%%%%%%%%%%%%%%%%%%%%%%%%%%%%%%%%%%%%%%%%%%%%%%%%%%%%%%%%%%

For quantum dots accommodating an odd number of electrons, a
suppression of charge-fluctuations in the Coulomb blockade regime
leads to a local spin-1/2 degree of freedom and, when temperature is
lowered through a characteristic Kondo-temperature $T_K$, the
Kondo-effect shows up as a zero-bias peak in the differential
conductance. In a dot with an even number of electrons, the two
electrons residing in the highest occupied level may either form a
singlet or promote one electron to the next level to form a triplet,
depending on the relative magnitude of the level splitting $\delta$
and the ferromagnetic intradot exchange energy $J$.  For $J>\delta$,
the triplet state prevails and gives rise to a zero-bias Kondo
peak~\cite{Sasaki00,Kogan03}, but when $\delta>J$ the singlet state
is lower in energy and no Kondo effect is expected in the linear
conductance. Nevertheless, spin-flip tunneling becomes viable when
the applied bias is large enough to induce transitions from singlet
to triplet state. Such inter-lead exchange-tunneling may give rise
to Kondo correlations and concomitant conductance peaks near
$V\sim\pm\delta/e$. However, since the tunneling involves excited
states with a rather limited life-time, the question remains to what
extend the coherence of such inelastic spin-flips and hence the
Kondo-effect is maintained?

In the context of a double-dot system, a qualitative description of
a finite-bias Kondo-effect, leaving out decoherence and
nonequilibrium effects, was given already in
Ref.\onlinecite{Kiselev03} and finite bias conductance peaks have
already been observed in carbon
nanotubes~\cite{Nygaard00,Liang02a,Babic04} as well as in GaAs
quantum dots~\cite{Jeong01,Kogan03,Zumbuhl04}. However, for lack of
a quantitative theory for this nonequilibrium resonance no
characterization of the phenomenon has yet been possible. As pointed
out in Refs.\onlinecite{Wegewijs01,Paaske04a,Golovach04}, a bias
induced population of the excited state (here the triplet), may
change a simple finite-bias cotunneling step into a cusp in the
nonlinear conductance. Therefore, in order to quantify the strength
of correlations involved in such a finite-bias conductance anomaly,
a proper nonequilibrium treatment will be necessary. As we
demonstrate below, the qualitative signature of Kondo-correlations
is a finite bias conductance peak which is markedly sharper than the
magnitude of the threshold bias.

%%%%%%%%%%%%%%%%%%%%%%%%%%%%%%%%%%%%%%%%%%%%%%%%%%%%%%%%%%%%%%%%%%%%%%%%%%%%%%%%%
%%%%%%%%%%%%%%%%%%%%%%%%%%%%%% Describing the experiment %%%%%%%%%%%%%%%%%%%%%%%%
%%%%%%%%%%%%%%%%%%%%%%%%%%%%%%%%%%%%%%%%%%%%%%%%%%%%%%%%%%%%%%%%%%%%%%%%%%%%%%%%%

%Fig1 was here
We have examined a quantum dot based on a single-wall carbon
nanotube (see Fig.~1a). Electron transport measurements of the
two-terminal differential conductance were carried out in a cryostat
with a base electron temperature of $T_{\rm el}\approx 80$~mK and a
magnetic field perpendicular to the nanotube axis. The low
temperature characteristics of the device are seen from the density
plot in Fig.~1b, showing ${\rm d}I/{\rm d}V$ as a function of
source-drain voltage $V$ and gate voltage $V_g$. The dominant blue
regions of low conductance are caused by Coulomb blockade (CB) while
the sloping white and red lines are edges of the CB diamonds, where
the blockade is overcome by the finite source-drain bias. Moreover,
white and red horizontal ridges of high conductance around zero bias
are seen. These ridges occur in an alternating manner, for every
second electron added to the nanotube dot. These are the Kondo
resonances induced by the finite electron spin $S=1/2$ existing for
an odd number $N$ of electrons where an unpaired electron is
localized on the tube. The zero bias resonances are absent for the
other regions (with even $N$) where the ground state spin is $S=0$.
Over most of the measured gate-voltage range (not all shown) the
diamond-plot exhibits a clear four-electron periodicity, consistent
with the consecutive filling of two non-degenerate sub-bands,
corresponding to the two different sub-lattices of the rolled up
graphene sheet, within each shell~\cite{Liang02a}. This
shell-filling scheme is illustrated in Fig.~1c.

%Fig 2 was here
We observe inelastic cotunneling features for all even $N$ at
$eV\sim\pm\Delta$ for a filled shell and $eV\sim\pm\delta$ for a
half-filled shell. Reading off the addition, and excitation energies
throughout the quartet near $V_g=-4.90$~V, shown in Fig.~1b, we can
estimate the relevant energy-scales within a constant interaction
model~\cite{Oreg00,Liang02a,Sapmaz05}. We deduce a charging energy
$E_{C}\approx 3.0$~meV, a level-spacing $\Delta\approx 4.6$~meV, a
subband mismatch $\delta\approx 1.5$~meV, and rather weak intradot
exchange, and intra-orbital Coulomb energies $J, dU < 0.05 \delta$.
The fact that $\delta > J$ is consistent with a singlet ground-state
for a half-filled shell, involving only the lower sub-band
(orbital), together with a triplet at excitation energy $\delta-J$
and another singlet at energy $\delta$. Notice that also the regions
with $N$ odd and doublet ground-state exhibit an inelastic resonance
at an energy close to $\delta$. We ascribe these to (possibly
Kondo-enhanced) transitions exciting the valence electron from
orbital 1 to orbital 2, but we shall not investigate these
resonances in detail. Focusing on the half-filled shell, i.e. $N=2$,
Fig.~2a shows the measured line-shape, $dI/dV$ vs. $V$, at
$V_{g}=-4.90$~V and for $T_{\rm el}$ ranging from 81~mK to 687~mK.
The conductance is highly asymmetric in bias-voltage and exhibits a
pronounced peak near $V\sim\delta/e$ which increases markedly when
lowering the temperature (cf. inset Fig.~2b).

%%%%%%%%%%%%%%%%%%%%%%%%%%%%%%%%%%%%%%%%%%%%%%%%%%%%%%%%%%%%%%%%%%%%%%%%%%%%%%%%%
%%%%%%%%%%%%%%%%%%%%%%%%%%%%%% Model and calculation %%%%%%%%%%%%%%%%%%%%%%%%%%%%
%%%%%%%%%%%%%%%%%%%%%%%%%%%%%%%%%%%%%%%%%%%%%%%%%%%%%%%%%%%%%%%%%%%%%%%%%%%%%%%%%

For $N=2$, we model the nanotube quantum dot by a two-orbital
Anderson impurity occupied by two electrons coupled to the leads via
four different tunneling amplitudes $t_{i\alpha}$, between orbitals
$i=1,2$ and leads $\alpha=L,R$. In terms of the conduction electron
density of states, $\nu_{F}$, which we assume to be equal for the
two leads, the tunneling induces a level-broadening
$\Gamma_{i}=2\pi\nu_{F}\sum_{\alpha}|t_{i\alpha}|^2$. In the
Kondo-regime, $\Gamma_{i}\ll E_{C}$, charge fluctuations are
strongly suppressed and the Anderson model becomes equivalent to a
Kondo-model incorporating second order super-exchange (cotunneling)
processes, like the one illustrated in Fig.~3a, in an effective
exchange-tunneling interaction (see Supplementary Information,
section 2). This effective model describes the possible transitions
between the five lowest-lying two-electron states on the nanotube
(see Fig.~3b). Since $J\ll\delta$, we neglect $J$ altogether and
assume that the four excited states are all degenerate with
excitation-energy $\delta$. The next excited state is a singlet
entirely within orbital 2 with excitation energy $2\delta$ (see
Supplementary Information, section 2). Although this state is merely
a factor of 2 higher in energy than the five lowest lying states, we
shall neglect this and all higher lying states altogether. Since
there are no spin-flip transitions connecting the ground-state
singlet to the excited singlet at energy $2\delta$, we expect this
omission to have a minute influence on the conductance for
$V<2\delta$.
%Fig 3 was here

In order to deal with the logarithmic singularities arising in
low-order perturbation theory within the Kondo model, it is
convenient to employ the so-called poor man's scaling
method~\cite{Anderson70}. As we have demonstrated
earlier~\cite{Rosch03a}, this method can be generalized to deal also
with non-degenerate spin-states, like a spin-$1/2$ in a magnetic
field, as well as with nonequilibrium systems where the local spin
degrees of freedom are out of thermal equilibrium with the
conduction electrons in the leads. The present problem is
complicated by the presence of the two finite energy scales, $eV$
and $\delta$, causing the renormalization of the couplings to
involve more than just scattering near the conduction electron
Fermi-surfaces. To accommodate for this fact, one must allow for
different renormalization at different energy scales, and the poor
man's scaling method should therefore be generalized to produce a
renormalization group (RG) flow of frequency dependent coupling
functions. The details of this method have been presented in
Refs.~\onlinecite{Rosch03a,Rosch05} and the specific RG-equations
for this problem are presented in section 3 of the Supplementary
Information.

These coupled nonlinear RG-equations do not lend themselves to
analytical solution but may be solved numerically with relatively
little effort. From this solution, we obtain renormalized
coupling-functions with logarithmic peaks at certain frequencies
determined by $eV$ and $\delta$~(see e.g. Fig.~3c). Since resonant
spin-flips take place only via the excited states at energy
$\delta$, Korringa-like spin-relaxation, via excitation of
particle-hole pairs in the leads, partly degrades the coherence
required for the Kondo-effect~\cite{Paaske04b}. Even at zero
temperature, the couplings therefore remain weak and increase
roughly as $1/\log(\sqrt{T^2+\Gamma^2}/T_{K})$, with $\Gamma$ being
the effective ($V$-dependent) spin-relaxation rate. Using the
parameters $t_{\alpha i}$ from the fit in Fig.~4, we calculate the
Kondo-temperature and the spin-relaxation rate and find that
$T_{K}=0.4$~mK and $\Gamma\approx 350$ mK (see Supplementary
Information, sections 3-5). This in turn implies a small parameter
$1/\ln[\delta/T_K]\approx 0.1$, which justifies our perturbative
approach. Notice that the smallness of $T_{K}$ corresponds roughly
to a mere 50\% reduction of the total hybridization to orbital 1
%($\sqrt{t_{L1}^{2}+t_{L2}^{2}}$)
as an electron is added to the nanotube. This tiny $T_{K}$ does not
show up directly in the $V$-dependence of the nonlinear conductance,
which is characterized instead by the excitation energy $\delta$ and
the spin-relaxation rate $\Gamma$.

%Fig 4 was here
Having determined the renormalized coupling functions, the
electrical current is finally calculated from the golden rule
expression:
\begin{multline}
I=\frac{2\pi e}{\hbar}\int_{-\infty}^{\infty}\!\!\!d\omega
\hspace*{-6mm}\sum_{\stackrel{\sigma,\sigma'=\uparrow,\downarrow}
{\gamma,\gamma'=s,-1,0,1,s'}}\hspace*{-6mm}
|g^{\gamma';\gamma}_{R,\sigma';L,\sigma}(\omega)|^{2} f(\omega-e
V/2) (1-f(\omega+\varepsilon_{\gamma'}-\varepsilon_{\gamma'}+e V/2))
n_{\gamma}\\-(V\to -V),
\end{multline}
where $n_{\gamma}$ are the nonequilibrium occupation numbers of the
five impurity states and
$g^{\gamma';\gamma}_{R,\sigma';L,\sigma}(\omega)$ is the
renormalized exchange-tunneling amplitude for transferring a
conduction electron from left to right lead and changing its spin
from $\sigma$ to $\sigma'$, while switching the impurity-state from
$\gamma$ to $\gamma'$. Figure~4 shows a comparison of this
calculation to the data, obtained by fitting the four tunneling
amplitudes $t_{i\alpha}$ determining the unrenormalized exchange
couplings. Notice that in practice the measured conductance at $V=0$
and at $|V|\gg\delta$, together with the asymmetry between positive
and negative bias, places strong constraints on these four
amplitudes. The resulting fit is highly satisfactory, with only
slight deviations at highest voltages.

To quantify the importance of Kondo-correlations, we compare also to
plain nonequilibrium cotunneling. This unrenormalized second order tunneling
mechanism clearly underestimates the conductance peak at positive
bias, and most severely so at the lowest temperature.
Qualitatively, simple nonequilibrium cotunneling gives rise to a cusp
in the conductance which is roughly as wide as the magnitude of the
threshold-bias itself. As evident from the data and analysis presented here,
nonequilibrium Kondo-correlations instead produce a conductance
anomaly which is sharper than the magnitude of the threshold,-- that
is a proper finite-bias peak.\\

\vspace*{4mm} \noindent {\bf\sf Acknowledgements}

\noindent {\scriptsize This research was supported by the Center for
Functional Nanostructures of the DFG (J.~P., P.~W.), by the European
Commission through project FP6-003673 CANEL of the IST Priority
(J.~P.), by ARO/ARDA (DAAD19-02-1-0039), NSF-NIRT (EIA-0210736)
(N.~M., C.~M.~M.) and the Danish Technical Research Council
(J.~N.).}

\vspace*{2mm} \noindent {\bf\sf Competing financial interests}

\noindent {\scriptsize The authors declare that they have no
competing financial interests.}

%%%%%%%%%%%%%%%%%%%%%%%%%%%%%%%%%%%%%%%%%%%%%%%%%%%%%%%%%%%%%%%%%%%%%%%%%%%%%%%%%
%%%%%%%%%%%%%%%%%%%%%%%%%%%%%% Figures %%%%%%%%%%%%%%%%%%%%%%%%%%%%%%%%%%%
%%%%%%%%%%%%%%%%%%%%%%%%%%%%%%%%%%%%%%%%%%%%%%%%%%%%%%%%%%%%%%%%%%%%%%%%%%%%%%%%%

\pagebreak

\begin{figure}[t]
\includegraphics[width=0.8\linewidth]{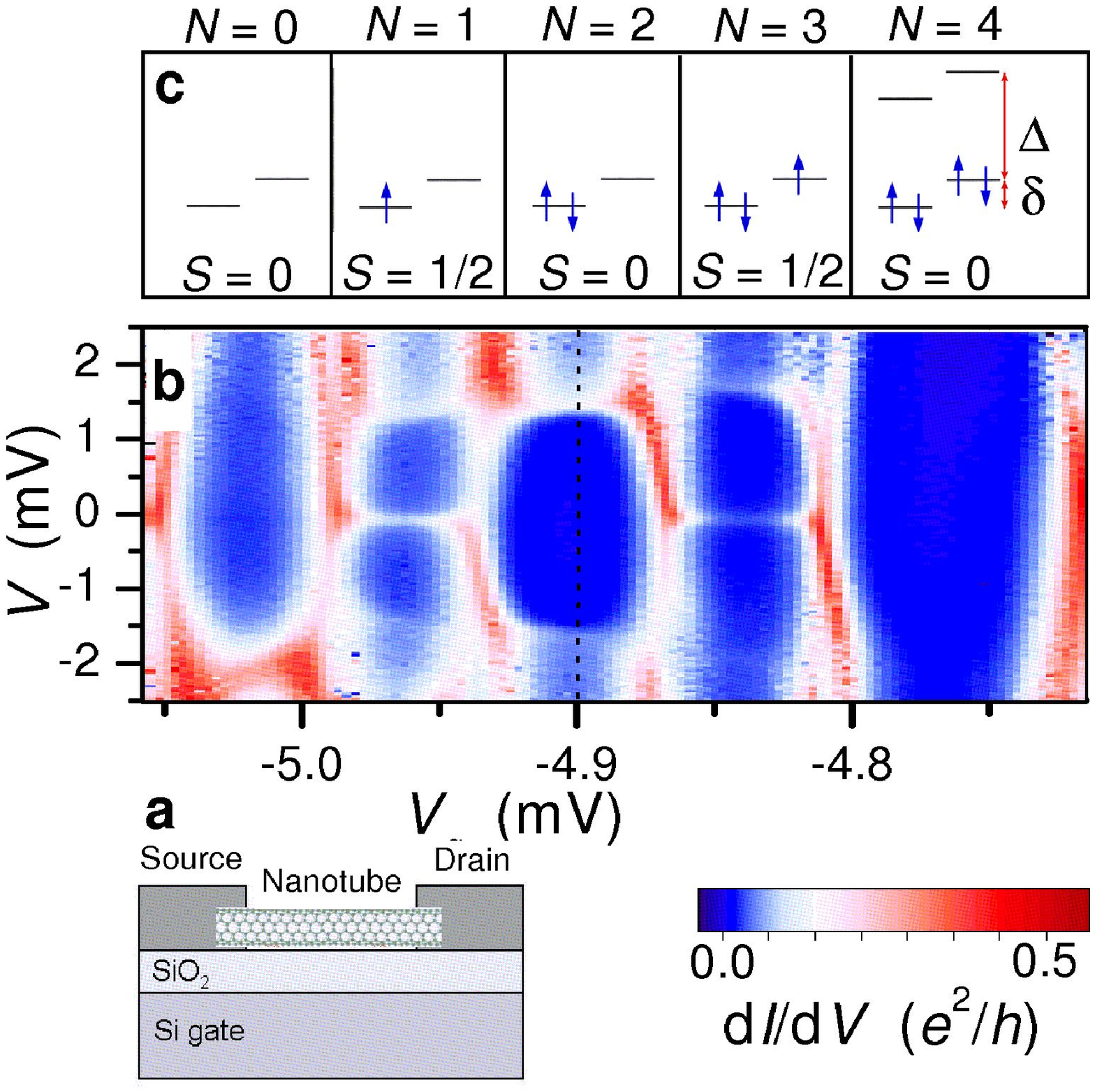}
\end{figure}
\noindent{\bf Figure 1 Experimental setup and shell-filling scheme
for a single-wall carbon nanotube.} {\bf a}, Schematic of the
nanotube device, comprising a single-wall carbon nanotube grown by
chemical vapor deposition on a SiO$_2$ substrate and contacted by
Cr/Au source and drain electrodes, spaced by 250~nm. Highly doped
silicon below the SiO$_2$ cap layer acted as a back gate electrode.
Room temperature measurements of conductance as a function of
back-gate voltage, $V_g$, indicate that the conducting nanotube is
metallic with a small gap outside the region considered here (cf.
Supplementary Information, section 1). {\bf b}, Density-plot of
$dI/dV$ as a function of $V$ and $V_{g}$ at $T_{\rm el}=81$~mK. {\bf
c}, Diagram illustrating the corresponding level-filling-scheme with
$N$ defined as the total number of electrons modulo 4.

\pagebreak

\begin{figure}[t]
\includegraphics[width=0.8\linewidth]{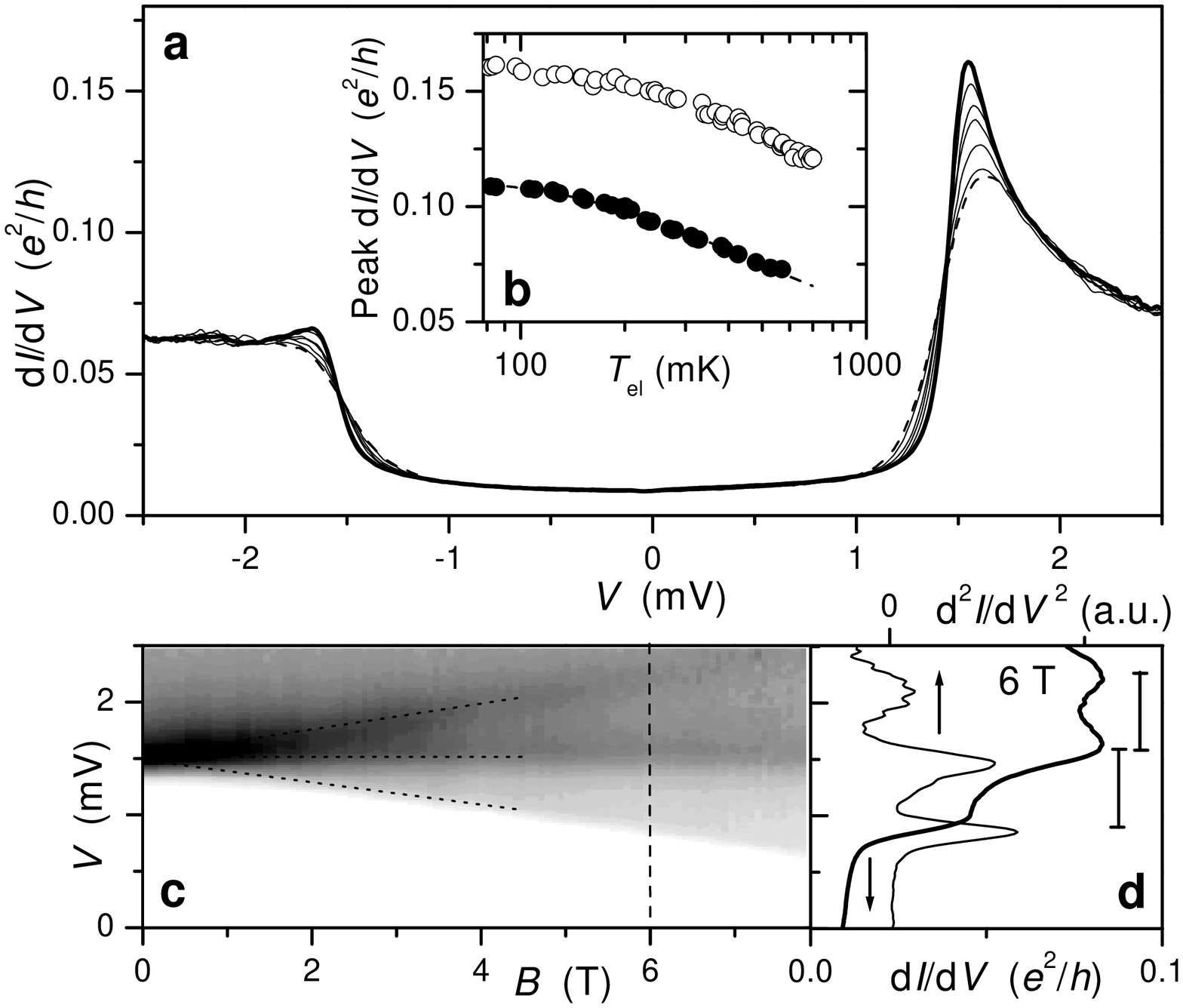}
\end{figure}
\noindent{\bf Figure 2 Temperature and magnetic field dependence of
the finite-bias resonance.} {\bf a}, $dI/dV$ as a function of $V$,
at $V_{g}=-4.90$~V (dashed cross section in Fig.~1b), taken at
$T_{\rm el}=$81~(thick), 199, 335, 388, 488, 614, 687~(dashed)~mK.
{\bf b}, Temperature dependence of the finite-bias conductance peak
in Fig.2a (open circles) and the neighboring zero-bias Kondo
resonance (solid circles) at $V_{g}=-4.96$~V (see Fig.1a). The open
circles are consistent with the saturation of a log-enhanced
finite-bias Kondo-peak as $T$ becomes smaller than the
spin-relaxation rate $\Gamma\approx 350$ mK, determined from the
parameters in Fig.~4. The solid circles follow the
NRG-interpolation-formula $a(1 +(2^{1/0.22}-1)(T/T_{K})^2)^{-0.22}$
with $a=0.11$ and $T_{K}=1.0$~K (dashed line). {\bf c}, $dI/dV$ vs.
$V$ and magnetic field $B$, at $V_{g}=-4.90$~V. {\bf d}, $dI/dV$ vs.
$V$ at $B=6$~T, corresponding to vertical cross section (dashed) in
Fig.2c. The $d^{2}I/dV^{2}$ trace (thin) underlines the presence of
three distinct peaks in $dI/dV$. The black bars each correspond to
$\Delta V=g\mu_{B}B/e$ with $g=2.0$ for nanotubes.

\pagebreak

\begin{figure}[t]
\includegraphics[width=0.8\linewidth]{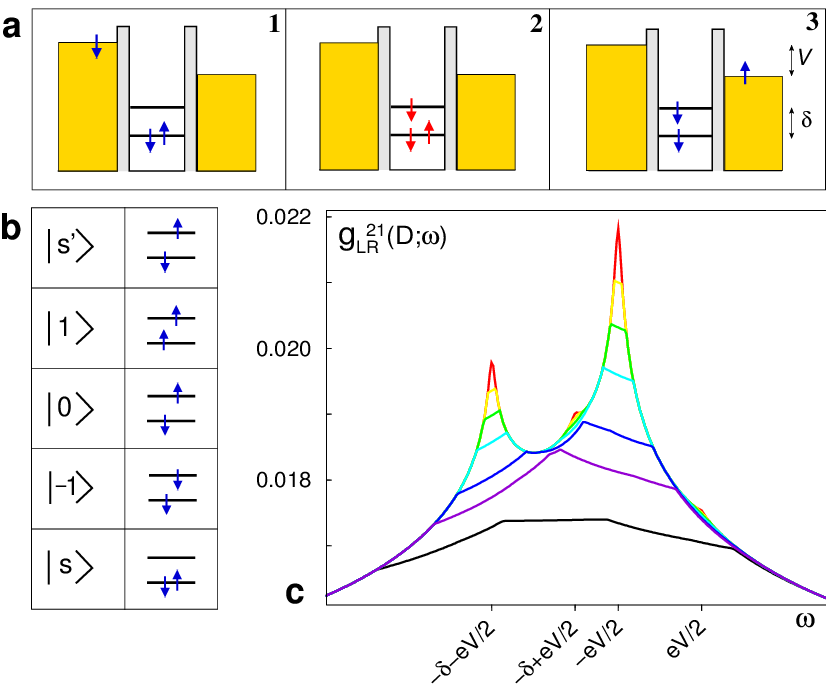}
\end{figure}
\noindent{\bf Figure 3 Schematic of the inelastic spin-exchange
underlying the Kondo-effect.} {\bf a}, Illustration of the
cotunneling mechanism giving rise to the effective
exchange-interaction in the Kondo model. Note that the virtual
intermediate state (red spins) in panel 2 is suppressed by a large
energy-denominator of the order of the electrostatic charging energy
of the nanotube. {\bf b}, Schematic of the five different low-energy
states retained in the effective Kondo-model: $|s\rangle$ is the
singlet ground-state, $|m\rangle$ (m=1,2,3) are the triplet
components with excitation energy $\delta-J\approx\delta$ and
$|s'\rangle$ is the corresponding singlet with excitation energy
$\delta$. Applying a magnetic field causes the triplet to split up
as seen in Fig.~2c,d. {\bf c}, Renormalized dimensionless
exchange-coupling vs. incoming conduction electron energy $\omega$,
measuring the amplitude for exchange-tunneling from right to left
lead while de-exciting an electron on the nanotube from orbital 2 to
1. Different colours encode the different stages of the RG-flow as
the bandwidth is reduced from $D_{0}$ to zero. Parameters as in
Fig.~4.

\pagebreak

\begin{figure}[t]
\includegraphics[width=0.8\linewidth]{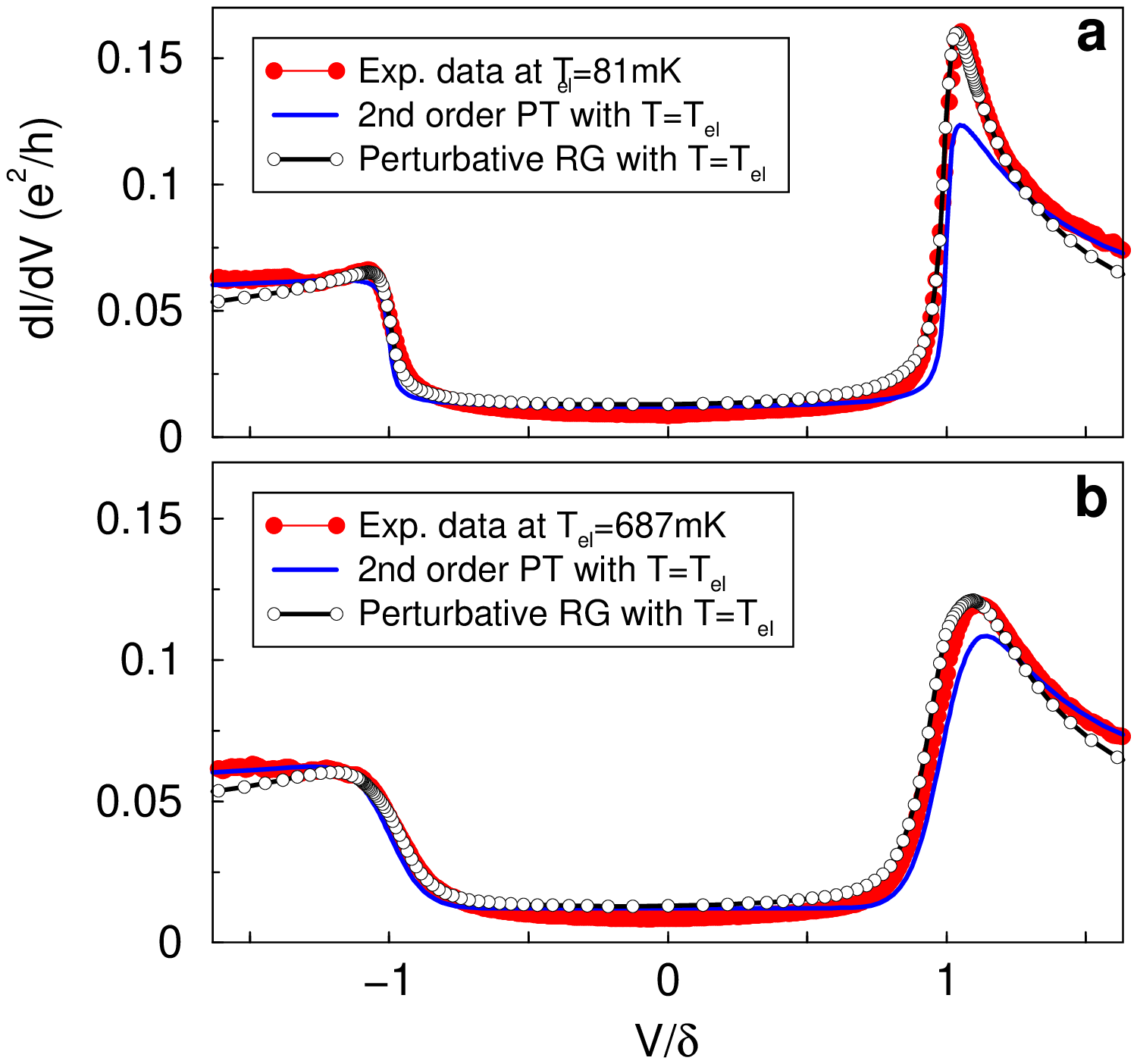}
\end{figure}
\noindent{\bf Figure 4 Fitting the nonlinear conductance by
perturbative RG calculation.} {\bf a}, Experimental data (red dots)
at the lowest temperature, fitted by the perturbative RG calculation
(white dots) with $\{t_{L1}, t_{L2}, t_{R1}, t_{R2}\} = \{0.024,
0.033,0.111, 0.083\}\sqrt{E_{C}/\nu_{F}}$ and $T=T_{\rm el}=81$~mK.
We use $J=0$ and an arbitrary bandwidth, $D_0=1$~eV, larger than all
other energy-scales. From the bias at which $dI/dV$ vs. $V$ has
maximum slope, we read off $\delta=1.515$~meV. The blue curve
represents simple cotunneling, i.e. second order perturbation
theory(PT) with tunneling-amplitudes $t_{i\alpha}$ fitted to the
data at $T=687$~mK (see panel b). {\bf b}, Comparing to the data
(red and white dots) at the highest temperatures, using same
parameters as in panel a, but with $T=T_{\rm el}=687$~mK. The blue
curve represents the best fit of second order perturbation theory
with $T=687$~mK.

%%%%%%%%%%%%%%%%%%%%%%%%%%%%%%%%%%%%%%%%%%%%%%%%%%%%%%%%%%%%%%%%%%%%%%%%%%%%%%%%%
%%%%%%%%%%%%%%%%%%%%%%%%%%%%%% Supplementary Information %%%%%%%%%%%%%%%%%%%%%%%%
%%%%%%%%%%%%%%%%%%%%%%%%%%%%%%%%%%%%%%%%%%%%%%%%%%%%%%%%%%%%%%%%%%%%%%%%%%%%%%%%%
\pagebreak

\begin{flushleft}
{\large\sf Nonequilibrium Singlet-Triplet Kondo Effect in Carbon
Nanotubes}

{\sf J. Paaske, A. Rosch, P, W\"{o}lfle, N. Mason, C. M. Marcus, J.
Nyg{\aa}rd.}
\end{flushleft}

\vspace*{5mm} \begin{flushleft} {\huge\bf\sf Supplementary
Discussion}
\end{flushleft}
\vspace*{10mm}

\noindent{\large\bf\sf 1. Nanotube transport experiments}

Our nanotubes were grown by chemical vapor deposition (CVD) on a
doped silicon substrate with a 400~nm oxide cap layer. Ferric iron
nitrate nanoparticles deposited from a solution in isopropyl alcohol
acted as catalyst for the CVD process, which was carried out in a
tube furnace by flowing methane  and hydrogen over the sample at
900~$^\circ$C \cite{hafner01}. This process yielded mostly
individual single-wall nanotubes as determined by atomic force
microscopy. The nanotubes were contacted by thermally evaporated
metal electrodes (35~nm Au on 4~nm Cr), spaced by 250~nm and
patterned by electron beam lithography. The two-terminal conductance
was measured using standard lock-in techniques with $\sim 5~\mu$V ac
excitation and voltage bias $V$ applied to the source with the drain
grounded through a low-impedance current amplifier. The effective
electron temperature can differ from the measured mixing chamber
temperature in the dilution refrigerator. A base electron
temperature of $T_{\rm el}\approx$ 80~mK was estimated from the
temperature dependence of device characteristics.

In the range of gate voltage considered in this paper, the room
temperature conductance of the device is around 1.8~$e^2/h$ and
independent of back-gate voltage $V_g$. At lower gate voltages, a
weak $V_g$ dependence at room temperature is seen and can be
attributed to a small band gap in the metallic nanotube induced by
perturbations.\cite{Minot04} The coupling decreases with $V_g$ and
at low $T$ the Coulomb blockade peaks disappear at $V_g\sim -2.5$~V,
i.e.\ around $\Delta N=40$ electrons away from the present $V_g$
range.\cite{herrero05}
\\

%%%%%%%%%%%%%%%%%%%%%%%%%%%%%%%%%%%%%%%%%%%%%%%%%%%%%%%%%%%%%%%%%%%%%
%%%%%%%%%%%%%%%%%%%%%%%%%%%%%% 2. %%%%%%%%%%%%%%%%%%%%%%%%%%%%%%%%%%%
%%%%%%%%%%%%%%%%%%%%%%%%%%%%%%%%%%%%%%%%%%%%%%%%%%%%%%%%%%%%%%%%%%%%%
\pagebreak\noindent{\large\bf\sf 2. Low energy two-particle states
and effective Kondo-Hamiltonian}

The relevant single-particle states of the nanotube carry spin
$\sigma=\uparrow, \downarrow$ and orbital index $i=1, 2$, and the
lowest lying two-particle states are denoted as follows:
\begin{eqnarray}
|\,s\rangle\,&=&\!\!(|\downarrow,\uparrow\rangle-|\uparrow,\downarrow\rangle)
\otimes|1,1\rangle/\sqrt{2}\\
|\!-\!1\rangle\,&=&\!\!|\downarrow,\downarrow\rangle
\otimes(|1,2\rangle-|2,1\rangle)/\sqrt{2}\\
|\,0\rangle\,&=&\!\!(|\downarrow,\uparrow\rangle
+|\uparrow,\downarrow\rangle)\otimes(|1,2\rangle-|2,1\rangle)/2\\
|\,1\rangle\,&=&\!\!|\uparrow,\uparrow\rangle\otimes(|1,2\rangle-|2,1\rangle)/\sqrt{2}\\
|\,s'\rangle\,&=&\!\!(|\downarrow,\uparrow\rangle
-|\uparrow,\downarrow\rangle)\otimes(|1,2\rangle+|2,1\rangle)/2\\
|\,s''\rangle\,&=&\!\!(|\downarrow,\uparrow\rangle
-|\uparrow,\downarrow\rangle)\otimes|2,2\rangle/\sqrt{2}
\end{eqnarray}
These states have energies $E_{s}$, $E_{-1,0,1}=E_{s}+\delta-J$,
$E_{s'}=E_{s}+\delta$ and $E_{s''}=E_{s}+2\delta$, and since
$J<\delta$, the singlet $|\,s\rangle$ is the ground-state. We
include only the five lowest lying states and neglect the highest
lying singlet, $|\,s''\rangle$, altogether. Within the Hilbert-space
of these five states, the effective Kondo-Hamiltonian takes the
form:
\begin{equation}
 H=\sum_{\stackrel{\stackrel{\scriptstyle {\bf k},\sigma}
                         {\scriptstyle i=1,2}}
                         {\scriptstyle \alpha=L,R}}
(\varepsilon_{{\bf k}}-\mu_{\alpha}) c^{\dagger}_{\alpha i {\bf
k}\sigma}c_{\alpha i {\bf k}\sigma}+H_{\rm int}%+H_{\rm dir}.
\nonumber
\end{equation}
where
\begin{eqnarray}
 H_{\rm int}&=&\frac{1}{2\nu_{F}}\!\!\!
\sum_{\stackrel{\stackrel{\scriptstyle {\bf k},{\bf
k}',\sigma,\sigma'}
                         {\scriptstyle i,j=1,2}}
                         {\scriptstyle \alpha,\alpha'=L,R}}
\!\!\!\{
\begin{array}{ccc}
   &  \\
   &  \\
   &  \\
\end{array}\hspace*{-4mm}
g^{ij}_{\alpha'\alpha}\left[ \delta_{ij}\,\vec{S}+
\tau^{3}_{ij}\,\vec{T}+
\tau^{1}_{ij}\vec{P}_{ij}\right]\cdot\vec{\tau}_{\sigma'\sigma}\nonumber\\
&&\hspace*{1.9cm}+p^{ij}_{\alpha'\alpha}\left[\delta_{ij}\,|s\rangle\langle
s|+ \frac{1}{2}\left(\tau^{+}_{ij}|s\rangle\langle s'|+
\tau^{-}_{ij}|s'\rangle\langle s|\right)\right]\delta_{\sigma'\sigma}\nonumber\\
&&\hspace*{1.9cm}+ q^{ij}_{\alpha'\alpha}\delta_{ij}
\!\!\!\!\sum_{m=-1,0,1,s'}|m\rangle\langle
m|\,\delta_{\sigma'\sigma}\,\, \} c^{\dagger}_{\alpha' i{\bf
k}'\sigma'} c_{\alpha j{\bf k}\sigma}\label{eq:hamilton}
\end{eqnarray}
with
\begin{equation}
\begin{tabular}{ll}
$\,\,g^{ii}_{\alpha'\alpha}=2\nu_{F}t_{i\alpha'}t^{\ast}_{i\alpha}/E_{C},$&
\hspace*{12mm}$g^{12}_{\alpha'\alpha}=(g^{21}_{\alpha\alpha'})^{\ast}
=2\sqrt{2}\nu_{F}t_{1\alpha'}t^{\ast}_{2\alpha}/E_{C},$ \\
$\!p^{ii}_{\alpha'\alpha}=\tau^{3}_{ii}g^{ii}_{\alpha'\alpha},$ &
\hspace*{10mm}$p^{12}_{\alpha'\alpha}=
(p^{21}_{\alpha\alpha'})^{\ast}=g^{12}_{\alpha'\alpha},$  \\
$q^{ij}_{\alpha'\alpha}=0.$   & $ $
\end{tabular}\label{eq:init}
\end{equation}
The vector of Pauli-matrices is denoted by  $\vec{\tau}_{ij}$ and
all terms in the interaction part have the form of spin, and orbital
exchange, except for the two terms in the last line proportional to
$\delta_{ij}\delta_{\sigma'\sigma}$ which are pure potential
scattering terms. Throughout, we use the convention
$A_{\pm}=A_{x}\pm i A_{y}$, for any vector-operator $\vec{A}$. The
Hamiltonian is expressed in terms of the two-particle
vector-operators
\begin{eqnarray}
S^{+}&=&(S^{-})^{\dagger}=
\sqrt{2}\left(\,|1\rangle\langle 0|+|0\rangle\langle-1|\,\right),\nonumber\\
S^{z}&=&
|1\rangle\langle 1|-|\!-\!1\rangle\langle-1|,\nonumber\\
P_{12}^{+}&=&(P_{21}^{-})^{\dagger}=
\sqrt{2}\,|1\rangle\langle s|,\nonumber\\
P_{12}^{-}&=&(P_{21}^{+})^{\dagger}=
-\sqrt{2}\,|\!-\!1\rangle\langle s|,\nonumber\\
P_{12}^{z}&=&(P_{21}^{z})^{\dagger}=
-|0\rangle\langle s|,\nonumber\\
T^{+}&=&(T^{-})^{\dagger}=
\sqrt{2}\left(\,-|1\rangle\langle s'|+|s'\rangle\langle-1|\,\right),\nonumber\\
T^{z}&=&|0\rangle\langle s'|+|s'\rangle\langle 0|\nonumber,
\end{eqnarray}
together with the scalar-operators
\begin{eqnarray}
|s\rangle\langle s|&=&\frac{1}{3}(P^{2}-\frac{1}{2}S^{2}),\nonumber\\
|s\rangle\langle s'|&=&(|s'\rangle\langle s|)^{\dagger}=
-\frac{1}{3}\vec{P}\cdot\vec{T},\nonumber\\
\sum_{m=-1,0,1,s'}|m\rangle\langle m|&=&\frac{1}{3}(T^{2}+S^{2}),
\end{eqnarray}
where $\vec{P}=\vec{P}_{12}+\vec{P}_{21}$. Defining also the
operator $M=[\vec{T},\vec{P}]/3i=i(|s\rangle\langle
s'|-|s'\rangle\langle s|)$, we note that the ten operators
$\{\vec{S}, \vec{T}, \vec{P}, M\}$ satisfy the commutation
relations:
\begin{equation}
\begin{tabular}{lll} $[S_{i},S_{j}]=i\varepsilon_{ijk}S_{k},$ &
\hspace*{10mm}$[S_{i},T_{j}]=i\varepsilon_{ijk}T_{k},$ &
\hspace*{10mm}$[M,T_{i}]=i P_{i},$  \\
$[T_{i},T_{j}]=i\varepsilon_{ijk}S_{k},$   &
\hspace*{10mm}$[S_{i},P_{j}]=i\varepsilon_{ijk}P_{k},$ &
\hspace*{10mm}$[M,P_{i}]=-i T_{i},$  \\
$[P_{i},P_{j}]=i\varepsilon_{ijk}S_{k},$   &
\hspace*{10mm}$[T_{i},P_{j}]=i\delta_{ij}M,$ &
\hspace*{10mm}$[M,S_{i}]=0.$
\end{tabular}
\end{equation}
which identifies them as generators of the Lie-algebra $SO(5)$.
Notice that $T^{2}+P^{2}+S^{2}+M^{2}=4$ is a Casimir-operator, i.e.
a constant of motion. The Hamiltonian (\ref{eq:hamilton}) is
manifestly invariant under spatial rotations, but it will posses
this abstract $SO(5)$-symmetry only when $\delta=J=0$ and
$t_{1\alpha}=t_{2\alpha}$, in which case one would expect to observe
a conventional zero-bias Kondo-peak, characterized by a
Kondo-temperature which is enhanced compared to the usual $SU(2)$
Kondo-effect. Notice that an $SO(5)$-Kondo-effect has been discussed
earlier in the context of a triple-quantum-dot
system\cite{Kuzmenko02}.

As pointed out for a double-dot system studied in
Ref.~\onlinecite{Kikoin01}, $\vec{S}$ and $\vec{P}$ generate
$SO(4)$, and it is the addition of the excited singlet $|s'\rangle$
which adds four new generators in the present problem. In a
double-dot system at even filling, the inter-dot tunneling breaks
the degeneracy between singlet and triplet states and an inelastic
cotunneling channel is generally available. A finite-bias
Kondo-resonance in such a double-dot system was suggested in
Ref.~\onlinecite{Kiselev03} and the current due to nonequilibrium
cotunneling was calculated in Ref.~\onlinecite{Golovach04}, but the
combined nonequilibrium Kondo-effect has not yet been examined. By
simply leaving out the excited singlet $|s'\rangle$, our present
calculation could readily be applied to this problem as well.\\

%%%%%%%%%%%%%%%%%%%%%%%%%%%%%%%%%%%%%%%%%%%%%%%%%%%%%%%%%%%%%%%%%%%%%
%%%%%%%%%%%%%%%%%%%%%%%%%%%%%% 3. %%%%%%%%%%%%%%%%%%%%%%%%%%%%%%%%%%%
%%%%%%%%%%%%%%%%%%%%%%%%%%%%%%%%%%%%%%%%%%%%%%%%%%%%%%%%%%%%%%%%%%%%%
\noindent{\large\bf\sf 3. Perturbative renormalization group
equations}

The (one-loop) perturbative renormalization group (RG) equations
satisfied by the frequency dependent couplings are established in
much the same way as explained earlier in Ref.~\onlinecite{Rosch05}.
Here, they take the following form:
\begin{eqnarray}
\frac{\partial g^{ii}_{\alpha'\alpha}(\omega)}{\partial\ln D}&=&
-\sum_{\alpha''}\left\{ g^{ii}_{\alpha'\alpha''}(\alpha''V/2)
g^{ii}_{\alpha''\alpha}(\alpha''V/2) \theta_{\omega-\alpha'' V/2}
\right.\nonumber\\&&\left.\hspace*{12mm} +\frac{1}{2}
g^{i\bar{i}}_{\alpha'\alpha''}(\alpha''V/2)
g^{\bar{i}i}_{\alpha''\alpha}(\alpha''V/2-\tau^{3}_{ii}\delta)
\theta_{\omega+\tau^{3}_{ii}\delta-\alpha'' V/2}
\right\},\label{eq:RGeqstart}
\\%&& \nonumber\\
\frac{\partial g^{i{\bar i}}_{\alpha'\alpha}(\omega)}{\partial\ln
D}&=&
-\sum_{\alpha''}\left\{\,\,\,\,\left[g^{ii}_{\alpha'\alpha''}(\alpha''V/2)+
\frac{1}{2}\left(p^{ii}_{\alpha'\alpha''}(\alpha''V/2)
-q^{ii}_{\alpha'\alpha''}(\alpha''V/2)\right)\right]
\right.\nonumber\\&&\hspace*{12mm}\times g^{i{\bar
i}}_{\alpha''\alpha}(\alpha''V/2+\tau^{3}_{ii}\delta)\theta_{\omega-\tau^{3}_{ii}\delta-\alpha''
V/2},
\nonumber\\
&&\hspace{12mm}+ \left[ g^{{\bar i}{\bar
i}}_{\alpha''\alpha}(\alpha''V/2)- \frac{1}{2}\left( p^{{\bar
i}{\bar i}}_{\alpha''\alpha}(\alpha''V/2)- q^{{\bar i}{\bar
i}}_{\alpha''\alpha}(\alpha''V/2)\right)
\right]\nonumber\\
&&\left.\hspace*{12mm}\times g^{i{\bar
i}}_{\alpha'\alpha''}(\alpha''V/2)\theta_{\omega-\alpha'' V/2}
\right\},
\\%&& \nonumber\\
\frac{\partial p^{ii}_{\alpha'\alpha}(\omega)}{\partial\ln D}&=&
-\frac{3}{2}\tau^{3}_{ii}\sum_{\alpha''} g^{i{\bar
i}}_{\alpha'\alpha''}(\alpha''V/2) g^{{\bar
i}i}_{\alpha''\alpha}(\alpha''V/2-\tau^{3}_{ii}\delta)
\theta_{\omega+\tau^{3}_{ii}\delta-\alpha'' V/2},
\\%&& \nonumber\\
\frac{\partial p^{i{\bar i}}_{\alpha'\alpha}(\omega)}{\partial\ln
D}&=& -\frac{1}{2}\sum_{\alpha''}\left\{\,\,
3g^{ii}_{\alpha'\alpha''}(\alpha''V/2)g^{i{\bar
i}}_{\alpha''\alpha}(\alpha''V/2+\tau^{3}_{ii}\delta)
\theta_{\omega-\tau^{3}_{ii}\delta-\alpha'' V/2}\right.\nonumber\\
&&\hspace*{14mm}+ 3g^{i{\bar
i}}_{\alpha'\alpha''}(\alpha''V/2)g^{{\bar i}{\bar
i}}_{\alpha''\alpha}(\alpha''V/2)
\theta_{\omega-\alpha'' V/2}\nonumber\\
&&\hspace*{14mm}+
\tau^{3}_{ii}p^{ii}_{\alpha'\alpha''}(\alpha''V/2)p^{i{\bar
i}}_{\alpha''\alpha}(\alpha''V/2+\tau^{3}_{ii}\delta)
\theta_{\omega-\tau^{3}_{ii}\delta-\alpha'' V/2}\nonumber\\
&&\hspace*{14mm}-\left. \tau^{3}_{ii}p^{i{\bar
i}}_{\alpha'\alpha''}(\alpha''V/2)p^{{\bar i}{\bar
i}}_{\alpha''\alpha}(\alpha''V/2) \theta_{\omega-\alpha''
V/2}\right\},
\\%&& \nonumber\\
\frac{\partial q^{ii}_{\alpha'\alpha}(\omega)}{\partial\ln D}&=&
\frac{1}{8}\tau^{3}_{ii}\sum_{\alpha''}\left\{\,\, 3g^{i{\bar
i}}_{\alpha'\alpha''}(\alpha''V/2)g^{{\bar
i}i}_{\alpha''\alpha}(\alpha''V/2-\tau^{3}_{ii}\delta)
\theta_{\omega+\tau^{3}_{ii}\delta-\alpha'' V/2}\right.\nonumber\\
&&\hspace*{10mm}+\left. p^{i{\bar
i}}_{\alpha'\alpha''}(\alpha''V/2)p^{{\bar
i}i}_{\alpha''\alpha}(\alpha''V/2-\tau^{3}_{ii}\delta)
\theta_{\omega+\tau^{3}_{ii}\delta-\alpha''
V/2}\right\},\label{eq:RGeqend}
\end{eqnarray}
with the shorthand notation $\theta_{x}=\theta(D-|x|)$, ${\bar 1}=2$
and ${\bar 2}=1$. For a given set of initial values (at scale
$D=D_{0}$), parametrized by the bare tunneling amplitudes
$t_{i\alpha}$ according to (\ref{eq:init}), these equations are
readily solved numerically for arbitrary $\omega$ and $D$. Taking
the limit of $D\to 0$, we obtain the renormalized coupling functions
used in the Golden Rule expression for the current.

As usual, when deriving the Kondo-model from an Anderson model, it
is convenient to introduce two angles, $(\cos\phi_{i},\sin\phi_{i})=
(t_{iL},t_{iR})/\sqrt{t_{iL}^{2}+t_{iR}^{2}}$, and parameterize the
exchange couplings as
\begin{equation}
\{g^{ij}_{\alpha'\alpha},p^{ij}_{\alpha'\alpha},q^{ij}_{\alpha'\alpha}\}=
\{g_{ij},p_{ij},q_{ij}\} \left(
\begin{array}{c}
\cos\phi_{i}\\
\sin\phi_{i}
\end{array}
\right)\left(
\begin{array}{cc}
\cos\phi_{j} & \sin\phi_{j}
\end{array}
%\begin{array}{cc}
%\cos(\phi_{i})\cos(\phi_{j}) & \cos(\phi_{i})\sin(\phi_{j}) \\
%\sin(\phi_{i})\cos(\phi_{j}) & \sin(\phi_{i})\sin(\phi_{j})
%\end{array}
\right)_{\alpha'\alpha}
\end{equation}
with initial conditions $g_{ij}=
2(\nu_{F}/E_{C})\sqrt{(1+\tau^{1}_{ij})(t_{iL}^{2}+t_{iR}^{2})(t_{jL}^{2}+t_{jR}^{2})}$,
$p_{ii}=\tau^{3}_{ii}g_{ii}$, $p_{12}=p_{21}=g_{12}$ and $q_{ij}=0$.
The L/R matrix-structure of the couplings is now exterior, in the
sense that the RG-equations are identical for every
$(\alpha',\alpha)$ component.

For $D\gg\delta, V, T, \omega$, the RG-equations simplify to
describe the flow of coupling constants $(g,p,q)_{ij}$ with no
dependence on frequency, $\omega$, nor lead-index, $\alpha=L, R$.
The difference in tunneling-strengths to the two orbitals still
plays an important role, and enters via the initial conditions. From
the RG-equations it is clear that all the different couplings
diverge at the same energy-scale and it is this scale which we
henceforth refer to as the Kondo-temperature, $T_{K}$ (see
Fig.~\ref{fig:fig1}).
\setcounter{figure}{4}\begin{figure}[t]
\includegraphics[width=0.5\linewidth]{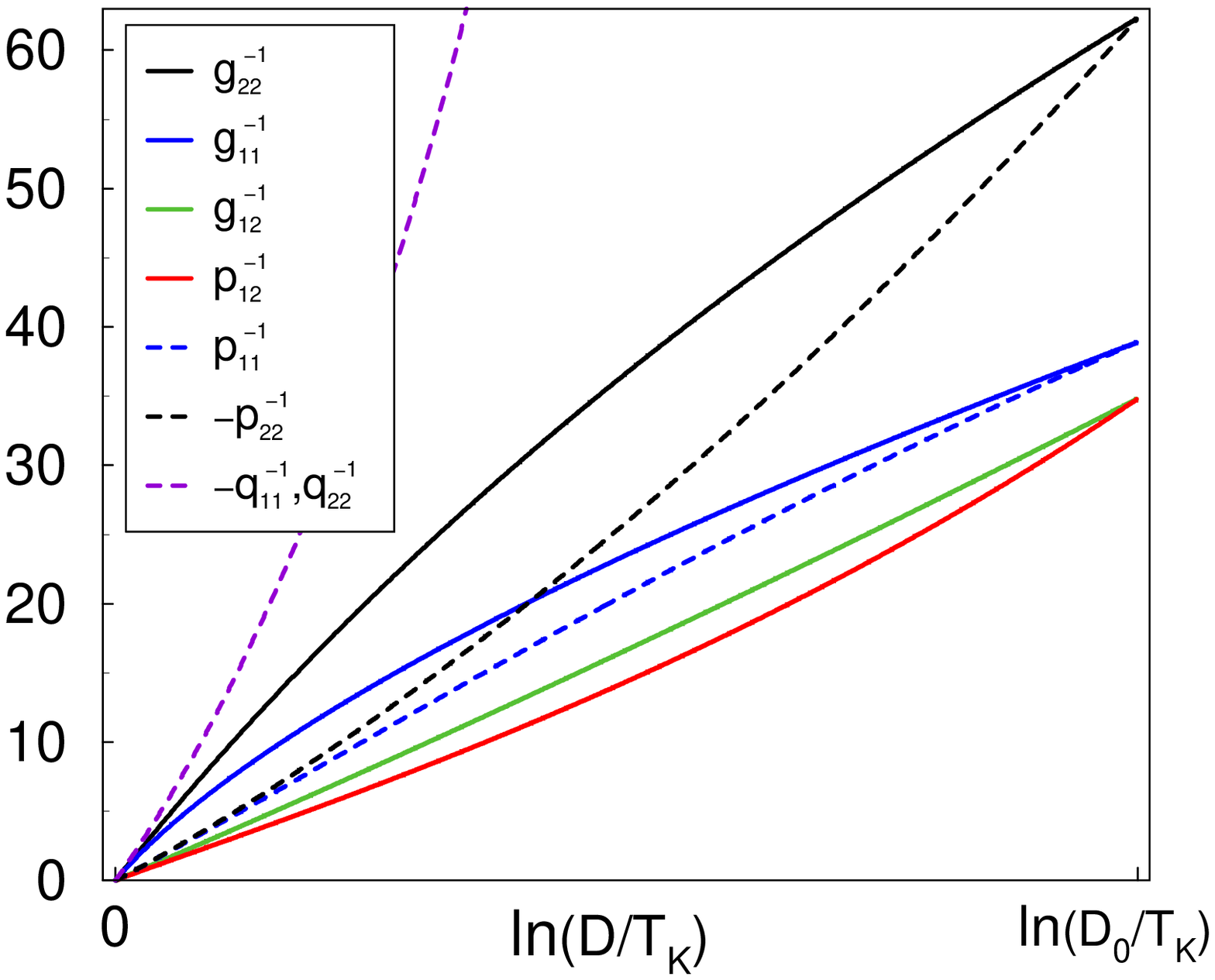}
\caption{\label{fig:fig1}{\bf Inverse couplings vs. bandwidth $D$.}
The inverse couplings all vanish at the same energy-scale defining
$T_{K}$. The couplings shown here are based on the same bare
parameters as were used for the low temperature fit in the main
paper: $\{t_{L1}, t_{L2}, t_{R1}, t_{R2}\} = \{0.024, 0.033,0.111,
0.083\}\sqrt{E_{C}/\nu_{F}}$, $J=0$ and $D_0=1$~eV, which implies
that $T_{K}\approx 0.4$~mK.}
\end{figure}
%Fig.~\ref{fig:fig1}).
%\begin{figure}[t]
%\includegraphics[width=0.5\linewidth]{FigSuppl1.eps}
%\caption{\label{fig:fig1}{\bf Inverse couplings vs. bandwidth $D$.}
%The inverse couplings all vanish at the same energy-scale defining
%$T_{K}$. The couplings shown here are based on the same bare
%parameters as were used for the low temperature fit in the main
%paper: $\{t_{L1}, t_{L2}, t_{R1}, t_{R2}\} = \{0.024, 0.033,0.111,
%0.083\}\sqrt{E_{C}/\nu_{F}}$, $J=0$ and $D_0=1$~eV, which implies
%that $T_{K}\approx 0.4$~mK.}
%\end{figure}
The numerical solution of these high-D RG-equations allows us to
study the dependence of $T_{K}$ on the three independent initial
values $g_{11}$, $g_{22}$ and $D_{0}$. This reveals the following
nearly perfect interpolation-formula for $T_{K}$:
\begin{equation}\label{eq:TK}
T_{K}\approx
D_{0}e^{-1/(g_{11}+g_{22}+1.6(g_{11}^{-1}+g_{22}^{-1})^{-1})},
\end{equation}
where 1.6 is a good approximation to a universal number describing
this particular Kondo-effect. For $g_{11}=g_{22}$ this reduces to
$T_{K}\approx D_{0}\exp(-0.36/g_{11})$, as found earlier in a study
of vertical quantum dots\cite{Pustilnik00}(cf. also
Refs.~\onlinecite{Eto02,Golovach03}).

Notice that in the standard Kondo-effect involving a zero-bias
conductance peak, the Kondo-temperature can be estimated directly
from the width of the conductance peak. In the present problem,
however, $T_{K}$ is of little physical significance insofar as the
non-linear conductance is characterized mainly by the
spin-relaxation rate $\Gamma$ (see sec. 5) and the subband-mismatch
$\delta$. Nevertheless, $T_{K}$ still encodes a scaling-property of
the RG-solutions and shows how our rather arbitrary choice of bare
bandwidth $D_{0}=1$~eV is linked to the values of the bare
couplings. In spite of the fact that the Kondo-temperature in this
problem is enhanced over that of the standard SU(2) symmetric
Kondo-model, our fitting parameters imply that $T_{K}\approx
0.4$~mK, which is much smaller than the value of $1.0$~K found for
the neighboring Coulomb-blockade valley having an odd number of
electrons on the tube (cf. Fig. 2b in main text). Estimating the
effective coupling in the neighboring valley from Eq.~(\ref{eq:TK})
with $g_{22}=0$, we find that this big difference in
Kondo-temperature corresponds to a mere reduction of the total
hybridization to orbital 1 ($\sqrt{t_{L1}^{2}+t_{R1}^{2}}$) by
roughly 50\% when changing from N=1 to N=2.
\\

%%%%%%%%%%%%%%%%%%%%%%%%%%%%%%%%%%%%%%%%%%%%%%%%%%%%%%%%%%%%%%%%%%%%%
%%%%%%%%%%%%%%%%%%%%%%%%%%%%%% 4. %%%%%%%%%%%%%%%%%%%%%%%%%%%%%%%%%%%
%%%%%%%%%%%%%%%%%%%%%%%%%%%%%%%%%%%%%%%%%%%%%%%%%%%%%%%%%%%%%%%%%%%%%
\noindent{\large\bf\sf 4. Nonequilibrium distribution functions}

Using the renormalized coupling-functions, we may calculate the
transition rates between the various two-particle states using the
Golden Rule expression
\begin{equation}
W_{\gamma'\gamma}(V, \delta,
T)=\frac{2\pi}{\hbar}\int_{-\infty}^{\infty}\!d\omega
\hspace*{-3mm}\sum_{\overset{\sigma,\sigma'=\uparrow,\downarrow}
{\alpha,\alpha'=L,R}}\hspace*{-3mm}%\sum_{\alpha,\alpha'=L,R}
|g^{\gamma';\gamma}_{\alpha',\sigma';\alpha,\sigma}(\omega)|^{2}
f(\omega-\mu_{\alpha})
(1-f(\omega+\varepsilon_{\gamma}-\varepsilon_{\gamma'}-\mu_{\alpha'})),
\end{equation}
where $f$ denotes the Fermi-function. The nonequilibrium
distribution functions for the two-particle states, $n_{\gamma}$,
are then found by solving the steady-state quantum Boltzmann
equation
\begin{equation}
\sum_{\gamma'}W_{\gamma\gamma'}n_{\gamma'}=
\sum_{\gamma'}W_{\gamma'\gamma}n_{\gamma},
\end{equation}
together with the constraint
$\sum_{\gamma=s,-1,0,1,s'}n_{\gamma}=1$, ensuring that the two
electrons in the half-filled shell on the nanotube occupy exactly
one of the five lowest two-particle states. These voltage-dependent
distribution functions are subsequently plugged into the Golden Rule
expression for the current.\\

%%%%%%%%%%%%%%%%%%%%%%%%%%%%%%%%%%%%%%%%%%%%%%%%%%%%%%%%%%%%%%%%%%%%%
%%%%%%%%%%%%%%%%%%%%%%%%%%%%%% 5. %%%%%%%%%%%%%%%%%%%%%%%%%%%%%%%%%%%
%%%%%%%%%%%%%%%%%%%%%%%%%%%%%%%%%%%%%%%%%%%%%%%%%%%%%%%%%%%%%%%%%%%%%
\noindent{\large\bf\sf 5. Spin-relaxation}

As demonstrated in Ref.~\onlinecite{Paaske04b}, a finite bias gives
rise to Korringa-like spin-relaxation via inter-lead particle-hole
excitations and for a single spin-1/2 the logarithmic singularities
were found to be cut off by the (V-dependent) spin-relaxation rates
$1/T_{1,2}$. To be more precise, it is the broadening of the
transverse dynamical susceptibility, $1/T_{2}$, which cuts off the
log-renormalization of spin-flip exchange couplings, whereas the
renormalization of non-spin-flip exchange couplings are contained by
the broadening of the longitudinal susceptibility, $1/T_{1}$.
Diagrammatically, these rates arise from a combination of both
self-energy, and vertex corrections to the spin-susceptibility
bubble and a simplification which simply omits the vertex
corrections will lead to serious mistakes.

In the present problem, the inter-orbital transition-rates (from
excited to ground-state) are strongly enhanced by the large
available phase-space. $W_{s,m}$ ($m=-1,0,1$) and $W_{s,s'}$ turn
out to be larger than all other transition-rates by roughly an order
of magnitude and the self-energy broadenings or line-widths of the
excited states are therefore much larger than that of the
ground-state. At least one of these large self-energy broadenings
will contribute to the cut-off in the log-renormalization of all
exchange couplings except one. $p_{ii}$ couples the ground-state to
itself and the relevant self-energy broadening is very weak. For
this particular coupling, vertex corrections will now play the
dominant role and they supply additional terms in the total
broadening involving the large $W_{s,m}$ and $W_{s,s'}$. All
log-singularities are therefore cut off by rates of the order of
these dominant line-widths of the excited states.

From the expression for the current (cf. main text Eq.(1)), it is
clear that the conductance peak at $V\sim\delta$ is mainly
determined by the inter-orbital exchange-couplings $g^{i\bar i}$ and
$p^{i\bar i}$. Furthermore, these couplings turn out to be larger
than the others and altogether the main influence of spin-relaxation
on the physical current is therefore via these inter-orbital
couplings. In other words, broadening all log-renormalization by a
single effective spin-relaxation rate estimated from an
inter-orbital susceptibility is expected to produce only minute
errors.

In terms of a generalized $ss'$-susceptibility, we determine the
physical spin-relaxation-rate $\Gamma_{s,s'}$  as the sum of
self-energy, and vertex corrections:
\begin{equation}
\Gamma_{s s'}= \Gamma^{s,s'}_{v}+\frac{1}{2}(\sum_{\gamma'\neq
s}W_{\gamma',s} +\sum_{\gamma'\neq s'}W_{\gamma',s'}),
\end{equation}
where $\Gamma^{s,s'}_{v}$ is the contribution from vertex
corrections. Just as the vertex-corrections to the transverse
spin-susceptibility for a single spin-1/2 include only non-spin-flip
processes\cite{Paaske04b}, the correction $\Gamma^{s,s'}_{v}$ picks
up only intra-orbital transition-rates and is therefore negligible
compared to the dominant term $W_{s,s'}$ coming from the self-energy
broadening of the excited state. $\Gamma_{s m}$ ($m=-1,0,1$) is
determined in a similar way and since these two different
inter-orbital relaxation rates turn out to be very close in
magnitude as well as in V-dependence we define a single effective
spin-relaxation rate as their average:
\begin{equation}
\Gamma=\frac{1}{2}(\Gamma_{s s'}+\Gamma_{s 0}).
\end{equation}

The spin-relaxation mechanism is incorporated in the RG-equations by
replacing $\theta_{x}$ by
$\theta(D-|\sqrt{x^{2}+T^{2}+\Gamma^{2}}|)$ in
Eqs.~(\ref{eq:RGeqstart}-\ref{eq:RGeqend}). Furthermore, all
Fermi-functions occurring in the transition-rates and the current
are effectively smeared by $\Gamma$, by replacing the
energy-conserving $\delta$-functions appearing in the Golden rule
expressions with Lorentzians of width $\Gamma$ (i.e.
spin-susceptibilitites ${\rm Im}[\chi^{R}]$). For example, we
evaluate the transition rate $W_{ST}$ as follows:
\begin{eqnarray}
W_{ST}(V, \delta,T) &=&\!\frac{2\pi}{\hbar}
\hspace*{-3mm}\sum_{\overset{\sigma,\sigma'=\uparrow,\downarrow}
{\alpha,\alpha'=L,R}}\!\!\int_{-\infty}^{\infty}\hspace*{-2mm}d\omega
\int_{-\infty}^{\infty}\hspace*{-2mm}d\varepsilon\,
|g^{\gamma';\gamma}_{\alpha',\sigma';\alpha,\sigma}(\omega)|^{2}
f(\omega-\mu_{\alpha}) (1-f(\omega+\varepsilon-\mu_{\alpha'})) {\rm
Im}[\chi^{R}_{ST}(\varepsilon)]\nonumber\\
&\approx &\!\frac{2\pi}{\hbar}
\hspace*{-3mm}\sum_{\overset{\sigma,\sigma'=\uparrow,\downarrow}
{\alpha,\alpha'=L,R}}\!\!\int_{-\infty}^{\infty}\hspace*{-2mm}d\omega
\int_{-\Lambda+\delta}^{\Lambda+\delta}\hspace*{-2mm}d\varepsilon\,
\left(\frac{1}{\mu_{\alpha}-\mu_{\alpha'}+\varepsilon}
\int_{\mu_{\alpha'}-\varepsilon}^{\mu_{\alpha}}\hspace*{-2mm}d\omega'\,
|g^{\gamma';\gamma}_{\alpha',\sigma';\alpha,\sigma}(\omega')|^{2}\right)\nonumber\\
&&\hspace*{45mm}\times
f(\omega-\mu_{\alpha})(1-f(\omega+\varepsilon-\mu_{\alpha'}))
\frac{\Gamma/\pi}{(\varepsilon-\delta)^2+\Gamma^2}\nonumber\\
&\approx &\!\frac{2\pi}{\hbar}
\hspace*{-3mm}\sum_{\overset{\sigma,\sigma'=\uparrow,\downarrow}
{\alpha,\alpha'=L,R}}\!\!
\int_{\mu_{\alpha'}-\delta}^{\mu_{\alpha}}\hspace*{-2mm}d\omega'\,
|g^{\gamma';\gamma}_{\alpha',\sigma';\alpha,\sigma}(\omega')|^{2}
\int_{-\Lambda}^{\Lambda}\hspace*{-2mm}d\varepsilon\,
(1+N(\varepsilon+\mu_{\alpha}-\mu_{\alpha'}+\delta))
\frac{\Gamma/\pi}{\varepsilon^2+\Gamma^2},
\end{eqnarray}
where $N$ is the Bose-function and
$\Lambda=\sqrt{\delta^2+J^2+V^2+T^2}$ is an ultra-violet cut-off on
the spin-susceptibility ensuring convergence of the integral over
$\varepsilon$. Notice that, in order to speed up the numerical
evaluation, we have replaced the square of the coupling-function by
its average over the window set by the Fermi-functions. The error
introduced by this approximation is estimated to be subleading in
the small parameter $1/\log(\delta/T_{K})$.

We find that $\Gamma(V=\delta)\approx 350$ mK (varying between 345
mK and 475 mK when changing $V$ over the measured range) and the
data clearly sample the full crossover from low to high temperatures
with $T_{\rm el}^{lowest}\approx 81$ mK $<\Gamma< 687$ mK $\approx
T_{\rm el}^{highest}$.

%%%%%%%%%%%%%%%%%%%%%%%%%%%%%%%%%%%%%%%%%%%%%%%%%%%%%%%%%%%%%%%%%%%%%%%%%%%%%%%%%
%%%%%%%%%%%%%%%%%%%%%%%%%%%%%% Bibliography for supporting online material %%%%%%
%%%%%%%%%%%%%%%%%%%%%%%%%%%%%%%%%%%%%%%%%%%%%%%%%%%%%%%%%%%%%%%%%%%%%%%%%%%%%%%%%

%%%%%%%%%%%%%%%%%%%%%%%%%%%%%%%%%%%%%%%%%%%%%%%%%%%%%%%%%%%%%%%%%%%%%%%%%%%%%%%%%
%%%%%%%%%%%%%%%%%%%%%%%%% Resetting size: small %%%%%%%%%%%%%%%%%%%%%%%%%%%%%%%%%
%%%%%%%%%%%%%%%%%%%%%%%%%%%%%%%%%%%%%%%%%%%%%%%%%%%%%%%%%%%%%%%%%%%%%%%%%%%%%%%%%
%\end{small}
%%%%%%%%%%%%%%%%%%%%%%%%%%%%%%%%%%%%%%%%%%%%%%%%%%%%%%%%%%%%%%%%%%%%%%%%%%%%%%%%%
%%%%%%%%%%%%%%%%%%%%%%%%%%%%%%%%%%%%%%%%%%%%%%%%%%%%%%%%%%%%%%%%%%%%%%%%%%%%%%%%%
%%%%%%%%%%%%%%%%%%%%%%%%%%%%%%%%%%%%%%%%%%%%%%%%%%%%%%%%%%%%%%%%%%%%%%%%%%%%%%%%%

\end{document}